# Ultrafast Neuromorphic Photonic Image Processing with a VCSEL Neuron

Joshua Robertson[*], Paul Kirkland, Juan Arturo Alanis, Matěj Hejda, Julián Bueno, Gaetano Di Caterina & Antonio Hurtado

**The ever-increasing demand for Artificial Intelligence (AI) systems is underlining a significant requirement for new, AI-optimised hardware. Neuromorphic (brain-like) processors are one highly-promising solution, with photonic-enabled realizations receiving increasing attention. Among these, approaches based upon Vertical Cavity Surface Emitting Lasers (VCSELs) are attracting interest given their favourable attributes and mature technology. Here, we demonstrate a hardware-friendly neuromorphic photonic spike processor, using a single VCSEL, for all-optical image edge-feature detection. This exploits the ability of a VCSEL-based photonic neuron to integrate temporally-encoded pixel data at high speed; and fire fast (100ps-long) optical spikes upon detecting desired image features. Furthermore, the photonic system is combined with a software-implemented spiking neural network yielding a full platform for complex image classification tasks. This work therefore highlights the potentials of VCSEL-based platforms for novel, ultrafast, all-optical neuromorphic processors interfacing with current computation and communication systems for use in future light-enabled AI and computer vision functionalities.**

## Introduction

A direct result of the vast uptake of internet-connected devices is the growing availability of data and the increasing demand for faster, more efficient data processing platforms. Electronic processing technologies have grown to alleviate some of this demand [1-3], demonstrating high computational throughput and enabling the development of novel systems for artificial intelligence (AI) [4]. However, less traditional computing approaches, such as those based on neuromorphic (brain-inspired) processing elements, have also risen in popularity [5,6]. These systems, that have thrived in electronics, have demonstrated highly parallel architectures and impressive decision-making capability. Nevertheless, like their more traditional processing counterparts, the performance increment of silicon-based platforms is becoming increasingly limited due to fundamental physical challenges in electronic technologies [7]. Crosstalk, parasitic capacitance, and Joule-heating each contribute to the limitation of the speed, bandwidth, footprint, and efficiency of electronic systems, in turn driving many researchers to investigate alternative platforms for future data processing systems.

       One such alternative platform is photonics. Photonic light-based systems boast features such as increased bandwidth, high energy efficiency, low cross talk and fast operation speeds, helping remedy some of the limitations posed to advancing electronics. Recently, investigations into photonic Artificial Neural Networks (ANNs) and neuromorphic systems have been on the rise. Optical devices, such as quantum resonant tunnelling (QRT) structures [8-10], optical modulators [11], phase-change materials (PCMs) [12] and semiconductor lasers (SLs) [13-17], to name a few, have all been investigated as candidates for novel neuromorphic photonic processing systems. Yet with the field still in its infancy, some investigations have already flourished into efforts to accelerate information processing in photonics with ANNs [18-20], and reservoir computing systems [21-22]. Similarly, Convolutional Neural Networks (CNNs), which have shown great success in the fields of image processing and

computer vision, are also seeing implementation on photonic platforms. Devices such as PCMs [23], micro-ring weighting banks [24] and modulators [25] have been proposed to improve the speed and efficiency of the computationally expensive convolution operations.

Like in classical ANNs, there are a number of processing tasks to which neuromorphic computing systems are particularly applicable, one being the classification and recognition of images and patterns. Software implemented neuromorphic Spiking Neural Networks (SNNs), similarly to CNNs, have demonstrated successful image processing functionalities. Such SNN systems have applied combinations of convolution and pooling techniques to extract image features [27], and have been used in conjunction with dynamic vision sensor (DVS) cameras [28] and single photon avalanche detectors (SPADs) [29] to perform image processing tasks. Further, reports have also shown that SNNs have lower energy operational requirements and produce lower latency than CNNs, due to only the active parts of the network requiring computation [30]. Therefore, SNNs made up of neuromorphic photonic devices (as reported theoretically in [31]), exhibit great potentials for the future development of fast and efficient image processing systems built directly on optical platforms.

Alongside these efforts, SLs have shown the ability to operate as high-speed artificial optical spiking neurons, emulating the functionality and spiking operation of biological neurons in the brain (see [26] and references therein for a review). Among SLs, vertical cavity surface emitting lasers (VCSELs) have attracted considerable attention given their unique attributes (e.g. high speed, ease to integrate in 2D arrays, low energy operation, high coupling efficiency to optical fibres, etc.). A number of neuronal functionalities have been recently demonstrated in VCSELs, including ultrafast spike activation and inhibition [32], networked spiking communication [33], spike rate encoding [34] and pattern recognition [35], to name a few. Furthermore, theoretical spin flip model (SFM) studies [36] indicate that VCSELs have the potential to present a pathway to unsupervised learning, via the neuron-inspired spike-timing dependent plasticity (STDP) technique. Also very recently, the authors have outlined the potentials of VCSEL neurons for image processing [37-38] and encoding functionalities [39]. However, in these early image processing reports, the VCSEL neurons acted as spike thresholding elements, requiring the image pixel information to be pre-processed offline.

This work overcomes this key challenge, demonstrating a photonic VCSEL-based hardware system and all-optical spiking convolutional operation for image processing functionalities. The proposed platform uses a single VCSEL acting as an artificial optical spiking neuron, in combination with time-division multiplexing to allow for an extremely hardware friendly platform. The integrate-and-fire capability of the spiking VCSEL neuron is exploited to simultaneously process multiple high-speed optical inputs (100ps-long), each encoding distinct image pixels to deliver spiking convolutional operation directly in the optical domain. This feature is used in this work to demonstrate all-optical edge-feature detection in (complex) source images at very high speed and with noise robust operation. We demonstrate successful edge feature detection for a variety of source images, including complex images (e.g. the logo of our Institute) as well as type-written and hand-written digits (5000 images from the MNIST database). Moreover, we show that the optical spiking outputs from the VCSEL-based neuromorphic photonic hardware system can be fed to software implemented SNNs to demonstrate successful hand-written digit image recognition. Additionally, our approach uses inexpensive and commercially-sourced VCSELs working at the key telecom wavelength of 1550 nm, hence permitting full compatibility with optical communication networks and data centre technologies.

## Results

### Spiking Image Edge Detection with a VCSEL Neuron.

In this work, we utilize the neuromorphic functionalities of VCSELs, in tandem with convolution, to perform image processing tasks. Specifically, we will tap into the input integration functionality of the VCSEL neuron that allows the device to temporally summate the energy contribution of multiple fast (100ps-long) input pulses occurring in quick succession [35]. Known in biological neuronal model as integration, this key processing functionality permits neurons to receive multiple input signals (through dendrite receptors) from different neurons in a network, integrate their combined intensity (in the neuron's body or soma) and fire spiking responses (once the threshold for spike firing is exceeded) that are communicated to neighbouring neurons (via its axon). Here, we capitalise on the optical integration capability of our VCSEL neuron, alongside its ability to threshold inputs and activate fast sub-ns spiking responses, to develop a novel optical approach to image edge detection [37].

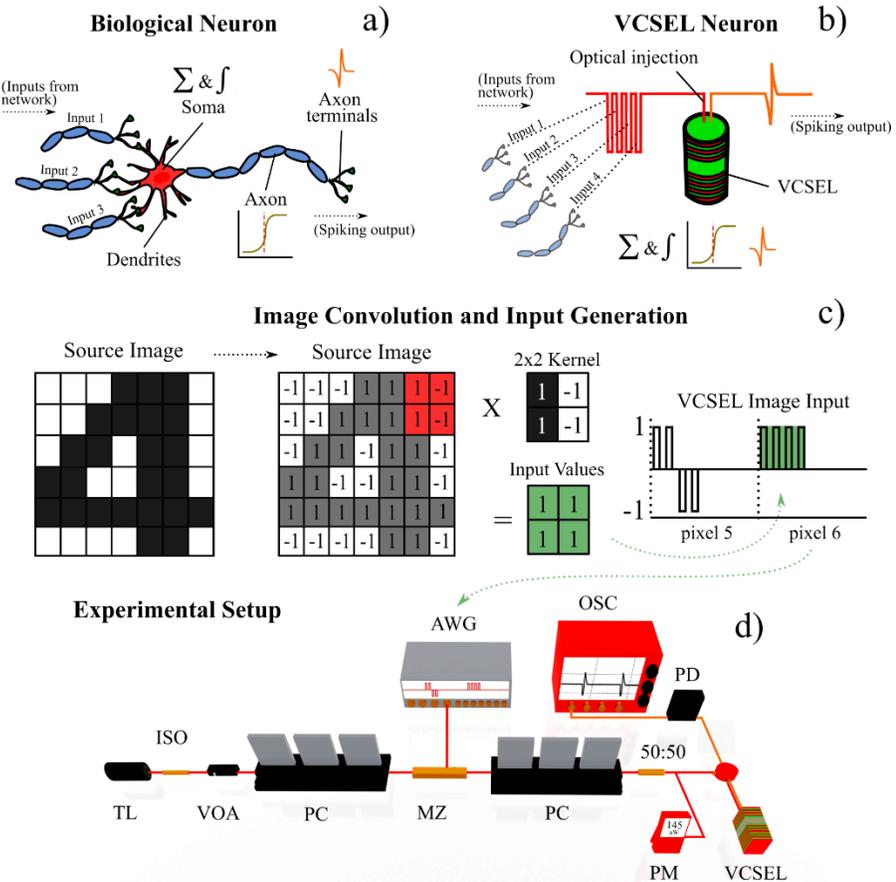

Fig. 1. Image processing technique with a spiking VCSEL neuron. Diagrams of a biological (a) and an artificial optical VCSEL neuron (b), illustrating the integration of multiple inputs and spike firing upon a threshold-exceeding input. Schematic of the image convolution and input generation process (c). Source images are converted into binary matrices, 1's (black) and -1's (white) and operated on by a 2x2 kernel operator creating a Hadamard (elementwise) product. The product is used for the generation of a return-to-zero (RZ) image input, where each pixel is given a configurable duration. The kernel operator is scanned across the image, time-multiplexing the inputs generated for each pixel. Experimental setup (d). The image input data is generated by an arbitrary waveform generator (AWG) and encoded, using a Mach Zehnder intensity modulator (MZ), into the light of a tuneable laser (TL). The injection light from the TL is directed into the VCSEL neuron using fibre-optic components; these include optical isolators (ISO), variable optical attenuators (VOA), polarisation controllers (PCs) and circulators (CIRC). The VCSEL neuron's response is analysed with a fast real time oscilloscope (OSC) after detection with a photodiode (PD).

In this new technique, image edge feature detection is performed according to the depiction illustrated in Fig. 1 (see methods section and supplementary information for full details). Fig. 2 shows experimental results on the edge-detection of a printed "Digit 4" 32x32 pixel source image. Here, 8 different (2x2) kernel operators are run sequentially. These kernels apply integer weights, 1 for black and -1 for white, to the image's pixel data. As 2x2 kernels are used here to process each pixel, 4 input values are required. These are encoded using 4 time-multiplexed 100 ps pulses. A total time of 3 ns per pixel was selected in the experiments and all pixel information is serialised before its injection into the VCSEL neuron.

First, two 'vertical' 2x2 kernels are applied to the source image triggering different spiking responses from the VCSEL neuron. These are captured using a real time oscilloscope and analysed using temporal maps, which de-multiplex the captured timeseries to form reconstructed images. Next, 3.0 ns segments (the configured pixel window) are sampled from the temporal maps to reveal which pixel inputs had activated spiking responses. A black pixel is plotted in a final image reconstruction when a spike activation (target feature) is detected. The convolution results for both vertical kernels are shown in final reconstruction of Figs. 2(c)-(d), revealing that the vertical edges of the "Digit 4" are identified by the VCSEL neuron (firing 100ps-long spikes). To achieve this result, the 'vertical' kernels yield bursts of 4 negative input pulses when a vertical edge feature is found in the image, and smaller input pulse bursts for all other pixel features (see supplementary information). The VCSEL neuron integrates the combined energy of these time-multiplexed bursts, only firing a spike when all four pulses are negative (the total energy goes beyond the spike firing threshold) to detect a vertical edge in the image. For all other cases, the VCSEL neuron remains quiescent as the combined energy of the encoded input burst does not exceed the spike activation threshold.

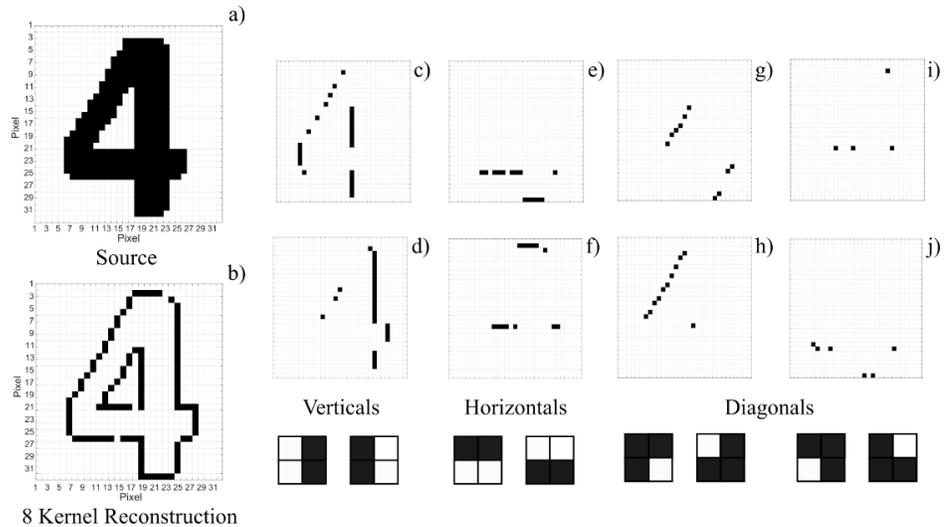

Fig. 2. Spike edge-feature detection of a black-and-white image with a VCSEL neuron. The 32x32 pixel source image (a) and reconstructed image (b) of a printed "Digit 4" after it's operated on by 8 different 2x2 kernels. Plots (c)-(j) depict individual results achieved for each of the 8 2x2 kernel operators utilised. These are shown in the insets below plots (c-j) and are respectively: two vertical (c)-(d), two horizontal (e)-(f), and four diagonal (g)-(j) kernel operators. Each black dot in plots (b-j) corresponds to a 100ps-long spike fired by the VCSEL neuron upon recognising a specific feature in the source image. The final 8 kernel reconstruction reveals all edges in the image.

In Figs. 2(e)-(f), we apply 2x2 'horizontal' kernel operators to detect the horizontal edge-features in the printed 'Digit 4' image. The 'horizontal' kernels again apply integer weights to the source image pixel data. The VCSEL neuron correctly reveals all horizontal features present in the image with the firing of fast sub-ns spikes. Additional kernel operators are applied to reveal the diagonal edges of the image. These are set with the following non-integer weights: 0.5, 0.75, 0.75 and -1 and the three rotations. The results of the diagonal integrate-and-fire convolutions are shown in Figs. 2(g)-(j), revealing successful detection of all diagonal features (Figs. 2(g)-(h)) as well as corner-features (Figs. 2(i)-(j)). Combining all eight experimental runs, we can form a single (8-kernel) reconstruction of the image revealing all edge information, as shown in Fig. 2(b). The technique presented here uses a single spiking VCSEL neuron combined with time-multiplexing and is therefore capable of providing all-optical edge detection functionality.

To further test the capability of our system we selected a source image of higher complexity, namely a 323x323 pixel RGB image of the University of Strathclyde's (UoS's) crest (Fig. 3(a)). This higher resolution image had three colour channels, red, green and blue. In this demonstration the green channel was selected before conversion into binarized black and white (Fig. 3(b)). We then used the same process and the same 8 kernel operators as in Fig. 2, to detect and reveal the image's edge features through the sub-ns spiking of the VCSEL neuron. The spiking time-series obtained with the 8 different 2x2 kernel operators are combined into the reconstructed image plotted in Fig. 3(c). The latter shows that all horizontal, vertical and diagonal lines were recognised; hence illustrating that neither image size nor complexity impede the operation of the photonic spiking VCSEL-based edge detection system. Importantly, an overall edge detection accuracy of 96.63% was achieved. A configurable pixel duration equal to 3.0 ns was set in all cases analysed in Figs. 2 and 3. Hence, our approach permits ultrafast image processing. For example, the time required to process the large 322 x 322 pixel UoS' crest image of Fig. 3 was equal to 3.0 ns x 322 x 322 = 311.1 μs per kernel operator. Importantly, we must note that the time-per-pixel could be still further reduced to values of just 1 ns (the spike refractory time of the VCSEL neuron), without any additional optimisation process, to achieve faster processing times.

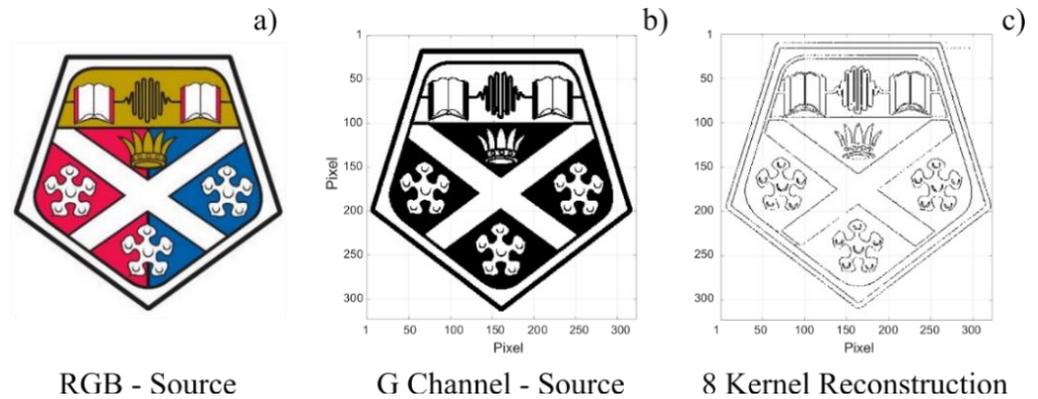

Fig. 3. Experimental edge detection of a complex, large size image with the VCSEL neuron. The 323x323 University of Strathclyde (UoS) crest source image (a) and its green (G-) colour channel component (b). The image is sequentially operated on by 8 2x2 kernels producing the reconstruction with all edge-features detected (c).

## Influence of Source Image Noise.

The performance of the edge-feature detecting VCSEL neuron system is next tested with 'noisy' images, see Fig. 4. Noise is implemented into the "Digit 4" source image via percentage variations of global pixel intensity. The same 8 2x2 kernel operators as in Figs. 2 and 3, are sequentially applied to the input image, to perform edge detection in a single experimental run. The performance of the system is measured by counting the number of successful detections. Fig. 4 reveals noise robust operation for the spiking VCSEL neuron system of this work, including successful operation under increasing levels of global pixel noise.

The results for globally increasing pixel noise are shown in Fig. 4(a)-(b). Initially, when no noise is added, the system responds revealing all edge-features in the image. Increasing the global noise to 5% and 10%, does not significantly affect performance, creating just 11% and 20.1% less successful activations for edge features. Increasing the global noise further to 15% causes the number of detections to drop overall (52.2% less activations). Finally, increasing image noise to 20% removes almost all feature detection (76.1% less activations) as now varying pixel intensities fail to match the kernel operator, creating input bursts incapable of spike activation. These results demonstrate that source image can be influenced by a global pixel intensity variation of up to 15% before kernel operators (configured for noiseless images) struggle to identify target features. Additional results, where intensity variations are introduced to the background of the printed "Digit 4" image are presented in the supplementary information.

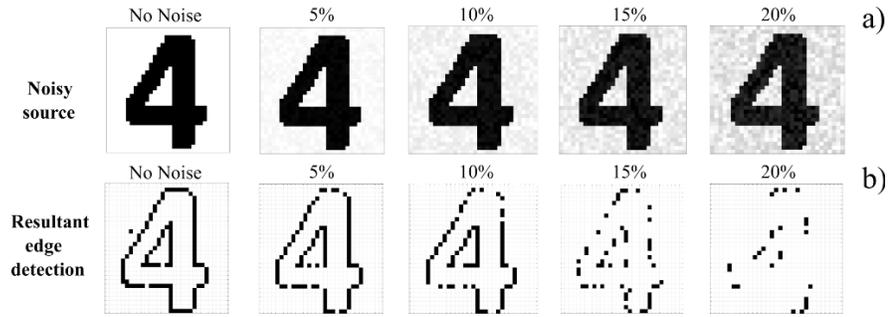

Fig. 4. Influence of source image noise on the edge-detection performance of the VCSEL neuron system. Noise is introduced to the printed "Digit 4" source image via the random variation of global pixel intensity (a)-(b). Global pixel intensity is varied randomly up to 20%.

**MNIST Handwritten Digit Edge Detection & Classification.** We proceed now to test our system with a large number of images from the MNIST handwritten digit database [40]. A total of 500 28x28 pixel images of each digit (0-9) are tested, with each image subject to convolution by 6 symmetrical 2x2 kernel operators. Time division multiplexing is used to sequentially perform each kernel operation on all 5000 images. An example of a source MNIST image and the resulting reconstructed image after convolution with 6 2x2 kernels are shown in Figs. 5(a)-(b). Experimental results for some selected examples of different digit images from the MNIST database are also provided in Fig. 5(c). These results show that symmetrical kernel operators and binary weights produce integrating bursts that successfully activate spike firing events, detecting edge-features for all 500 consecutive images (processed in a single experimental run) for each of the 10 digits.

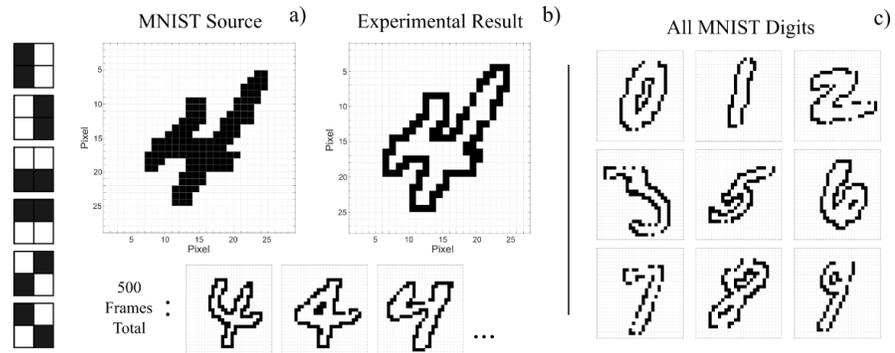

Fig. 5. MNIST handwritten-digit edge detection with an integrate-and-fire VCSEL neuron. Six 2x2 symmetrical kernels (2 vertical, 2 horizontal, 2 diagonal) we applied to 500 images of each digit from the MNIST dataset. All kernels are sequentially applied to the source image (a) which, when analysed and combined, produce reconstructions (b) that reveal all edge information. Each set of MNIST digits (c) were processed separately.

The single VCSEL neuron system completes a run of 500 MNIST hand-written digit images in just 6.56 ms (3 ns x 27 x 27 x 6 kernels x 500 images). This high-speed operation of 13.12 μs per MNIST image, already permitting processing >76,000 images per second, could still be significantly improved without any additional optimisation stage, just by reducing the pixel duration from its current set value of 3 ns to levels closer to the spiking refractory period of the VCSEL neuron (equal to approx. 1 ns for our commercially-sourced devices [34]); this would allow for a direct improvement of processing speed by a factor of 3. However, it should be noted that the spiking refractory period of the VCSEL neuron could be even further reduced (to sub-ns levels) with bespoke device designs permitting higher spiking rates. Similarly, the work reported here demonstrates an extremely hardware-friendly system using a single VCSEL for time-multiplexed convolutional operations, yielding output results in the form of fast (sub-ns long) spikes directly in the optical domain. We note that network architectures with multiple VCSEL neurons operating simultaneously would also enable parallel computation, further increasing the speed of the system at the expense of increasing system's complexity.

This work also bridges the gap between spike processing photonic hardware and software systems. To that end, we combine the photonic VCSEL-based spike image edge-feature detection system with a software implemented SNN to deliver full MNIST handwritten digit (HWD) image classification. To do so, the experimental photonic spiking data produced by the spiking VCSEL neuron system, containing the edge-feature information of 5000 MNIST handwritten digit images (see Fig. 5), was passed to a software SNN, which utilised hierarchical spike feature extraction and contained two further convolutional layers and a 10 feature fully-connected SoftMax layer for classification. The structure of the SNN is described further in the supplementary information. Fig. 6 (a) shows examples of outputs from the photonic edge detection system (MNIST digit 6 image) that become the inputs for the software SNN, which subsequently provides the SNN classification probability (Fig. 6(b)).

The classification performance of the combined photonic spiking hardware and software SNN system is shown in Fig. 6 (c), with a very high average accuracy of 96.1%. The performance of this of hierarchical feature extraction SNN proved more impressive than that of a fully connected layer alone, which, when tested, produced an accuracy of 92.2%, highlighting the benefits of additional convolution layers in computer vision. The accuracy of this SNN system compares very well with recent results reported in other neuromorphic photonic systems based upon phase change materials [23], where an accuracy of 96.1% was achieved using 4 edge detection kernels and a fully connected layer. However, unlike reported convolutional neural network demonstrations, the software implemented SNN provides a reduction in computational requirements, using around 10% of the computational power required for CNN operation, despite using multiple time steps to make a classification.

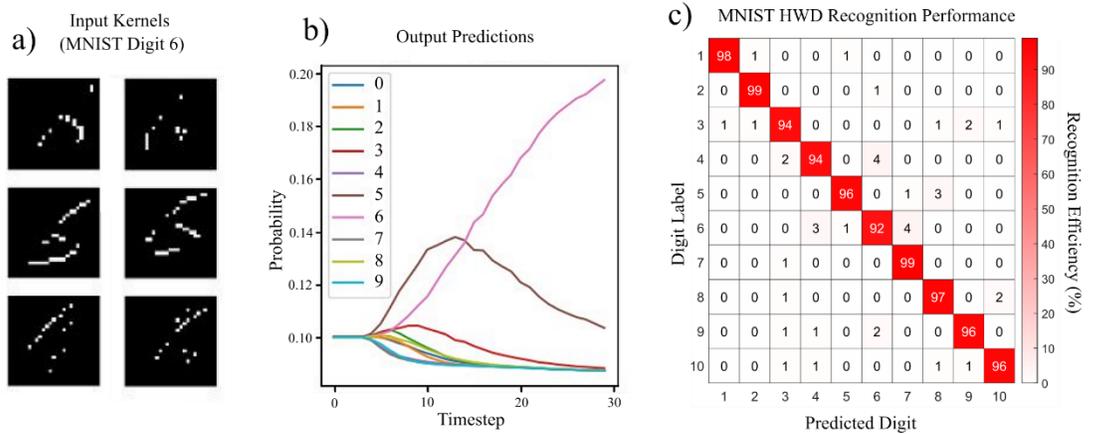

Fig. 6. MNIST handwritten-digit classification with the hardware-software SNN. Six input feature maps, previously produced by the spiking VCSEL neuron (a) are introduced to the software implemented SNN and the resulting classification probability is plotted over multiple training cycles (b). Results for an MNIST 'Digit 6' are plotted in (a) & (b). A confusion matrix shows the task performance for each class of digit across the 5000 MNIST HWD images tested. The SNN produces an overall average performance of 96.1%.

The successful edge detection result coupled with the high classification accuracy of the SNN suggest that photonic hardware and software implemented SNNs can be combined for spike-based image processing systems. Further, the successful results indicate that a fully photonic SNN network, based on VCSEL neuron convolutional layers, could be implemented for image processing systems, where future challenges lie in increasing the size of the kernel operators. This characteristic has also been investigated in this work in theory. Specifically, we analysed numerically the operation of our VCSEL neuron system with larger 3x3 kernel operators. Full details on the numerical model, equations and parameters, as well as all numerically calculated results, are provided in Supplementary Information. Our theoretical investigations showed the successful detection of target image features, as well as full edge-feature detection for larger 3x3 kernel operators. This demonstrates that the VCSEL neuron can successfully integrate larger bursts of input pulses (9 for the case of 3x3 kernel operators) with ultrashort temporal durations and separations between consecutive pulses (within the fast integration time-window of the device), that go beyond the capabilities of our experimental setup. The numerical findings therefore show that it could be possible to implement multiple layers of VCSEL neurons, such as those currently implemented by the software in this work (Fig. 6), towards the realisation of future VCSEL-based spike-based photonic image processing platforms.

## Discussion

We demonstrate a neuromorphic photonic system for image processing using a single VCSEL as an artificial spiking neuron. The system benefits from high speed operation (using 100ps-long inputs) and hardware-friendly implementation, relying on just a single VCSEL device and time division multiplexing. The proposed technique utilises the temporal input integration, thresholding, and spike firing capabilities of the VCSEL neuron, to perform all-optical spiking convolution on complex source images with a variety of kernel operators. This capability is used to demonstrate all-optical neuromorphic image edge-feature detection with a VCSEL neuron. Using streams of optical input pulses, we showed that consecutive 2x2 kernel operators and images can be run with a hardware-friendly single VCSEL platform, outputting fast neuromorphic spiking events for the detection of target edge features. Moreover, our approach showed very good robustness to image noise. We demonstrated that the system can successfully process 5000 images from the benchmark MNIST handwritten digit database. We showed that 500 images (per digit) can be processed in a single experimental run within 6.56 ms (at 13.12 μs per image) using commercial devices and components at telecom wavelengths, without any specific VCSEL optimisation stages. Additionally, combining the experimental photonic spiking outputs from the VCSEL neuron with a software-implemented SNN, we achieved a mean image classification accuracy of 96.1%, highlighting the potential of our approach for high-speed, low-energy spike-based image processing. Finally, we demonstrated theoretically that the operation of the VCSEL neuron with larger dimension (e.g. 3x3) kernels for more complex image feature extraction functionalities is also possible. This implies that VCSEL neurons have the potential to implement further convolution tasks, whether it be SNN layers (such as those in our software implemented SNN) or in recognition systems that target specific features. Overall, we believe that artificial spiking VCSEL neurons show high potential for future high speed, low energy, and hardware friendly neuromorphic photonic platforms for image processing with a fast telecom-compatible spiking representation.

## Methods

### Experimental Setup.
The fibre-based optical injection setup used for image processing (edge-feature detection) with an artificial optical spiking VCSEL is shown in Fig. 1. Light from a tuneable laser source (TL) is passed through an optical isolator to prevent reflections before entering a variable optical attenuator (VOA) to control optical injection power. A polarisation controller (PC) is used to maximize the performance of the 10 GHz Mach Zehnder intensity modulator (MZ), responsible for the optical encoding of the image input. Image inputs are generated by a 12 GSa/s, 5 GHz arbitrary waveform generator (AWG, Keysight M8190a) and amplified using an electrical amplifier before being fed into the MZ modulator. A second PC is then used to set the final polarisation of the optical injection. A coupler is used to monitor the optical injection power via a power meter (PM), and an optical circulator is used to inject the signal into the VCSEL neuron. Temporal analysis of the VCSEL neuron's output is performed using a 9 GHz photodetector (PD - Thorlabs PDA8GS) and an 8 GHz, 20 GSa/s real-time oscilloscope (OSC - Rohde & Schwarz RTP). In this work the VCSEL is driven with a bias current of 4.0 mA ($I_{th}$ = 0.83 mA) and is temperature stabilized at 293 K. The VCSEL device exhibited single transverse mode lasing with two orthogonal polarisation modes (device characterisation provided in supplementary information). Injection polarization was matched to that of the dominant (parallel) mode of the device and was made with a negative frequency detuning from the peak, inducing injection locking. The encoded inputs were configured to produce short (100ps-long) drops or raises around the mean optical power of the injected signal (145 µW). When injected input pulses integrate sufficiently, the injection power drops below the locking threshold, inducing a locking/unlocking transition into a dynamical regime of excitable spiking dynamics. This mechanism is responsible for the neuronal functionality of the VCSEL neuron, allowing it to trigger fast sub-nanosecond (approx. 100ps-long) spiking events in response to target edge features, directly in the optical domain.

**Image Edge Detection.** Image edge detection is performed according to Fig. 1. The pixel intensities of the source images are converted into integers ('1' for black and '-1' for white). This is achieved either by averaging across RGB colour channels, or by selecting a specific colour channel, converting it to greyscale and using a configurable pixel intensity threshold to binarize the pixels intensities. During the convolution process, kernel operators apply weights to customizable regions of the source image, producing Hadamard products. The local pattern descriptor identifies the region of the source image that requires sampling for kernel operation. In this work, the local pattern descriptor is a square M x N pixel area (highlighted in red in Fig. 1 (c)) with the anchor pixel present at M=N=1. The local pattern descriptor has a (M+1) x (N+1) range of 2x2 pixels. No image padding was used during convolution, hence the final dimensions of the convolved image were reduced by 1. In this work, we demonstrate the in-system integration of multiple data inputs by the VCSEL neuron, thus performing the pooling of the Hadamard product. To achieve this, we encode the weighted pixel values into a (return-to-zero) RZ signal, where each value is assigned an individual pulse. Each encoded input pulse has an amplitude corresponding to its Hadamard product value, and a duration of ~100 ps FWHM. A peak-to-peak separation of ~150 ps is used between input pulses, with zero padding also added to fill the pixel to a configurable window. In this work a pixel window of 3.0 ns was selected, higher than the refractory period of the spiking dynamics in the VCSEL neuron (approx. 1 ns long), allowing each pixel to independently activate spiking responses. This encoding scheme makes use of time-division multiplexing to encode the Hadamard product into a sequence of input pulses and encode multiple convolution operations sequentially into a single device. The activation threshold, governed by the injection power and frequency detuning, is required to be set such that only inputs burst associated with image target features trigger activations.

**Influence of Noise on Edge Detection Performance.** To implement global noise all pixel intensity was varied randomly according to the configurable noise percentage (%). The source image was implemented with 0%, 5%, 10%, 15% and 20% global noise. The weights of the kernel operators had to be altered such that the activation threshold was consistent for all integrating bursts across the 8 kernel operations. The maximum integrated input was therefore normalized by adjusting the vertical and horizontal kernel weights to the non-integer value of 0.75, such that 0.75+0.75+0.75+0.75 = 3. Finally, the convolutions of all 5 noisy images were combined into a single image input, such that the activation threshold was consistent across all tested images.

## Data availability

All data underpinning this publication are openly available from the University of Strathclyde KnowledgeBase at https://doi.org/10.15129/cfc1e947-9afe-40fd-bb4b-c7e271a77941.

## Acknowledgments

The authors acknowledge support from the UKRI Turing AI Acceleration Fellowships Programme (EP/V025198/1), the US Office of Naval Research Global (Grant ONRG-NICOP-N62909-18-1-2027), the European Commission (Grant 828841-ChipAI-H2020-FETOPEN-2018-2020), the EPSRC Doctoral Training Partnership (EP/N509760), and Leonardo MW Ltd through the Leonardo Lectureship at Strathclyde.

## Author contributions

J.R, M.H, J.A.A, & J.B carried out the experimental work under the supervision of A.H. P.K obtained classification results with the software SNN under the supervision of G.D. All authors helped identify the presented work, discussed the results, and contributed to the writing of the manuscript.

## Competing interests

The authors declare no competing interests.

# Ultrafast Neuromorphic Photonic Image Processing with a VCSEL Neuron

*Joshua Robertson\*, Paul Kirkland, Juan Arturo Alanis, Matěj Hejda, Julián Bueno, Gaetano Di Caterina & Antonio Hurtado*

## Supplementary Information

## Table of Contents

**Vertical Cavity Surface Emitting Laser (VCSEL) Characterisation.** A single mode vertical-cavity surface-emitting laser (VCSEL) was used to create a photonic spiking processing system for image edge-feature detection. The VCSEL used in this work was a commercially available device operating at the telecom wavelength of 1550 nm. The device was tested and characterised prior to the operation demonstrated in the manuscript. Fig. S1 plots the VCSEL's output power and lasing spectra at different bias currents when temperature stabilised at 293 K. Fig. S1a plots the L-I curve of the device measured at 293 K, showing a threshold current of $I_{th} = 0.83$ mA. When biased above threshold the device delivers continuous wave (CW) light emission. Fig. S1b shows the red shift of lasing spectra with increasing applied bias current. The VCSEL used in this work was a single-longitudinal and single-transverse mode laser source, which had two coexisting linear and orthogonally-polarised modes, referred to here as the orthogonally-polarised ($\lambda_x$) and parallel-polarised ($\lambda_y$) modes. The device operated with a dominant parallel $\lambda_y$ mode and did not exhibit bias-induced polarisation switching across the measured operating parameters. At 4.0 mA the peak wavelength of the dominant $\lambda_y$ mode was found at 1286.97 nm and the peak of the subsidiary $\lambda_x$ mode at 1287.11 nm. The corresponding wavelength separation between the two orthogonally-polarised modes for this device was measured as 0.136 nm (24.7 GHz). In this work injection of the image data input was made into the parallel $\lambda_y$ mode of the device with injection polarisation matching the dominant mode of the VCSEL.

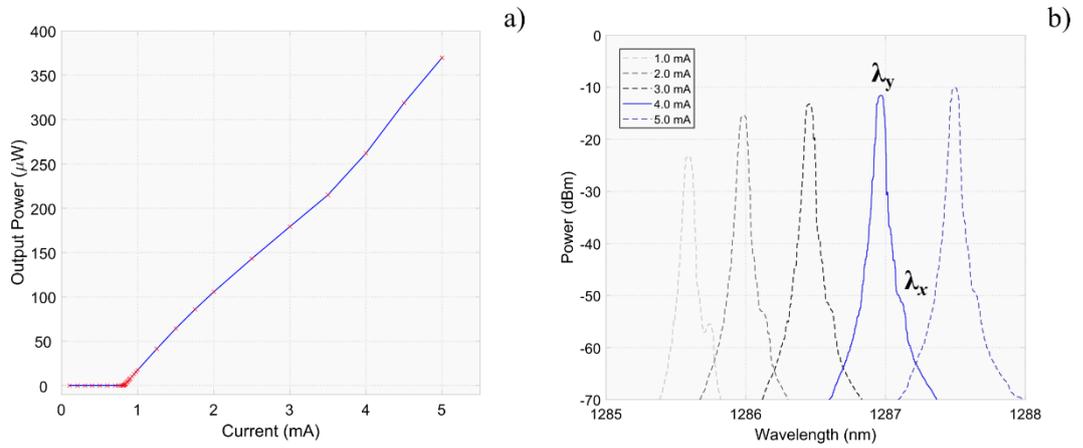

a)

b)

Fig. S1. Characterisation of the VCSEL. The output power was measured against bias current to determine the threshold of the device (a). The threshold current was equal to 0.83 mA. The lasing spectra of the device was recorded for increasing bias currents using an optical spectrum analyser (b). At the operating current of 4.0 mA the device demonstrated two orthogonal polarisation modes ($\lambda_y$ and $\lambda_x$) with peak wavelength of 1286.97 nm and 1287.11 nm respectively.

**Image Inputs and Optical Integration.** The photonic VCSEL neuron is responsible for the integration of optical inputs, the thresholding of their total contribution and the activation of fast spiking dynamics. Here, as described in Fig. 1 in the main article, image inputs are created by time-multiplexing the Hadamard product values of convolution operations, forming return-to-zero (RZ) encoded waveforms with bursts of positive and negative pulses. The RZ waveforms, or image inputs, are subsequently generated in an arbitrary waveform generator and encoded into the optical intensity of a tunable laser for injection into the VCSEL neuron. When the encoded injection enters the device, the inputs are integrated and the VCSEL neuron responds with fast spike events when its threshold for spike firing is exceeded. In this report the input pulses are generated at a rate of 12 GSa/s (1 sample per input), creating ~100 ps long pulses. The input pulses are grouped into fast ~700 ps bursts at the beginning of a configurable pixel window (equal to 3 ns in this work). Following the 3 ns window the pulse burst of the next pixel in the convolution process is generated. The generated image input is encoded into the optical injection entering the VCSEL neuron such that positive inputs produce negative intensity drops. Fig. S2a shows an example of encoded optical injection. The optical injection corresponds to the convolution of pixels 14-19 in row 13 of the printed "Digit 4" image with a 2x2 vertical kernel.

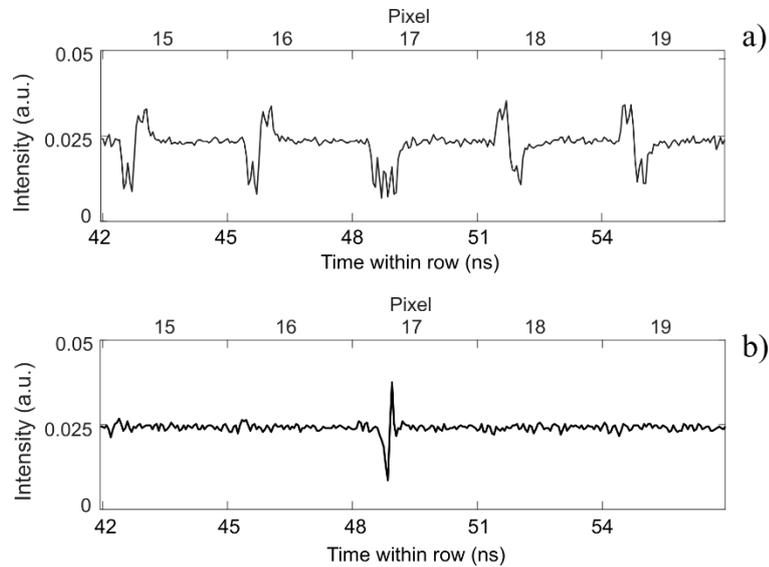

Fig. S2. Input-encoded injection and the subsequent VCSEL neuron response. The encoded optical injection, incident on the VCSEL neuron, is shown for the the convolution of pixels 14-19 in row 13 of the printed "Digit 4" with a 2x2 vertical kernel (a). The VCSEL neuron responds to the incoming signal by firing a spiking event for a target feature detection (pixel 17) (b).

As expected, the encoded optical injection takes the form of sequential bursts of positive and negative pulses. When a mismatch between kernel and local pattern descriptor (selected pixels) is present, combinations of positive and negative pulses occur within the same burst, as in the case of pixels 15,16,18 and 19 (see Fig. S2a). These inputs, when integrated by the VCSEL neuron, produce a low energy contribution to the activation threshold, subsequently failing to activate a spiking response in the recorded VCSEL neuron output shown in Fig. S2b. In contrast, when a kernel matches the local pattern descriptor, the injection is encoded with 4 negative inputs, as shown in pixel 17. Once injected into the neuron, the 4 negative inputs combine to produce a change in injection power that crosses the threshold for spike activation. This subsequently triggers an excitable spike at the output of the VCSEL neuron, as shown in Fig S2b. Following the successful or unsuccessful activation of spiking responses the system is reset during the remainder of the 3 ns window. This approach to input encoding and optical integration is used throughout this work including each demonstration of edge detection with the VCSEL neuron and the initial photonic hardware layer of the SNN used for the classification of MNIST hand-written digit (HWD) images.

**Background Noise Variation.** As described in the main article, the edge detection performance of the system was tested on the printed "Digit 4" image for increasing source image noise. In addition to testing increasing global pixel noise, we have tested performance on source images with increasing background noise. To implement background noise in source images the intensity of each white pixel was varied randomly between its current value (-1) and a percentage (%) of the maximum intensity value (1). For a background noise of 100% white pixels could vary randomly between white (-1) and black (1). The source image was implemented with 0%, 20%, 40%, 60% and 80% background noise. Figs. S3(a) and 4(b) analyse the case of increasing background pixel noise, revealing good overall resilience to noise. Initially, when no noise is added, the system responded revealing all edge-features in the image. Increasing the background noise to 20% (Figs. 4(a)-(b)), did not dramatically affect performance, decreasing slightly (25.9% reduction) the number of successful detections (firing of sub-ns spikes). Increasing the background noise to 40% and 60% we observed the number of detections reduced overall (57.7% and 79.8% less successful activations). Despite this, the image became unrecognisable, and performance further decreased as background noise reached 80%. These results show that without altering the threshold of the system, target features can still be recognised up to 40% background noise.

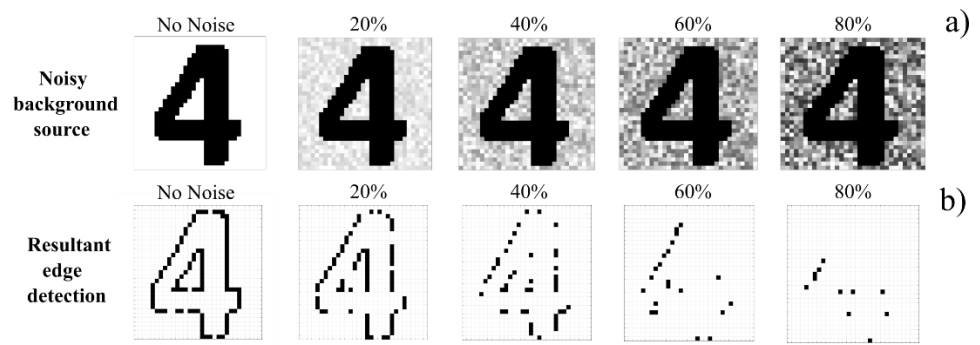

Fig. S3. Influence of source image noise on the edge-detection performance of the VCSEL-Neuron system. Noise was introduced to the printed "Digit 4" source image via the random variation of background-pixel intensity (a)-(b), and the random variation of global pixel intensity (c)-(d). Background pixels had their intensity values randomly varied by up to 80%, global pixel intensity was varied up to 20%.

**Spiking Neural Network (SNN) Structure.** A software implemented spiking neural network (SNN) is incorporated alongside a photonic VCSEL neuron spiking edge detection system to achieve the classification of MNIST handwritten digit images. As described in the main article, the optical spiking data generated by the experimental VCSEL neuron is fed to the software implemented SNN. The structure of the complete system is shown in Fig. S4.

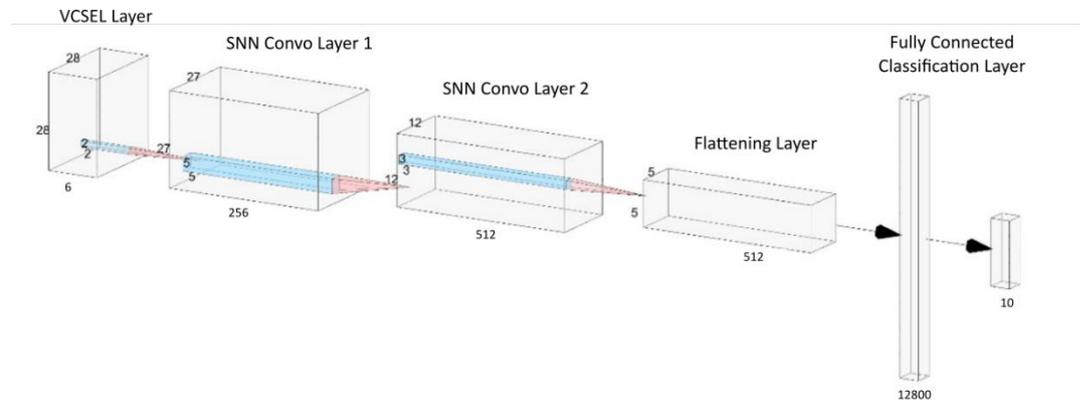

Fig. S4. The structure of Spiking Neural Network used in this work in combination with the VCSEL neuron photonic system for the classification of MNIST images. The first layer of the SNN is the experimental VCSEL neuron which performs spiking convolution with 2x2 kernel operators, creating a photonic spiking data stream. The software implemented SNN is fed the experimental data, subsequently performing convolution in 2 additional convolutional layers. The output of the second convolutional layer is flattened and passed to a 10 feature fully connected layer for classification.

The first layer of the SNN system is built experimentally with a single VCSEL neuron. This layer uses 6 2x2 kernel operators to yield edge-feature detection outputs in the form of fast optical spikes (as shown in Fig. 5 of the main article). The photonic spiking data is detected (with a >9.5 GHz bandwidth amplified photodetector) and recorded using a fast real-time oscilloscope. The recorded spiking time series are then fed into the software-implemented spiking neural network. The software implemented SNN made use hierarchical feature extraction and hence implemented a further two convolutional layers. The first SNN convolution layer created 256 feature maps using 5x5 kernel operators with a stride of 2x2, reducing the dimension of the feature maps to 12x12. The second convolution layer created 512 feature maps using 3x3 kernel operators again with a stride of 2x2, further reducing the dimension of the feature maps to 5x5. A flattening layer was used to prepare the network for the final 10 feature fully connected SoftMax layer which provided output probabilities for the final classification of the MNIST HWD image. The classification of 5000 MNIST HWD images were tested in this work and the recognition efficiencies are provided in Fig. 6 of the main article.

The network was constructed with the help of the Tensorflow and Keras libraries before being converted to a leaky integrate and fire system using the Nengo DL libraries. This method allowed backpropagation through gradient descent to be used to train the network weights prior to the conversion into a SNN. The completion of the training before SNN conversion also granted the use of well-established software training tools. In this work multiple training time steps were used before the final performance of the network was measured, as demonstrated in Fig. 6 of the main article. Overall, the method of converting a CNN to a SNN helped simplify implementation and training but did however influence the performance of the system. The average performance of the MNIST HWD classification reduced from 97.9% to 96.1% following the conversion to an SNN. This drop is performance is a trade off with the improvement to computational requirements, with the SNN using around 10% of that required by the CNN.

**Theoretical Analysis of Larger Kernel Operation.** To investigate further the potential of the neuromorphic image processing system of this work, convolution with larger (3x3) kernel operators was simulated using the Spin-Flip Model (SFM). The SFM equations (SI 1-5) were modified to include additional optical injection terms and were solved using the fourth order Runge-Kutta method:

$$\frac{dE_{x,y}}{dt} = -\left(k \pm \gamma_a\right)E_{x,y} - i\left(k\alpha \pm \gamma_p\right)E_{x,y}$$
$$+ k\left(1+i\alpha\right)\left(NE_{x,y} \pm inE_{x,y}\right) + k_{inj}E_{inj}\left(t\right)e^{i\Delta\omega_x t} + F_{x,y} \tag{SI 1}$$

$$\frac{dN}{dt} = -\gamma_N\left[N\left(1+\left|E_x\right|^2 + \left|E_y\right|^2\right) - \mu + in\left(E_y E_x^* - E_x E_y^*\right)\right] \tag{SI 2}$$

$$\frac{dn}{dt} = -\gamma_s n - \gamma_N\left[n\left(\left|E_x\right|^2 - \left|E_y\right|^2\right) + iN\left(E_y E_x^* - E_x E_y^*\right)\right] \tag{SI 3}$$

$$F_x = -\sqrt{\frac{\beta_{sp}\gamma_N}{2}}\left(\sqrt{N+n}\,\xi_1 + \sqrt{N-n}\,\xi_2\right) \tag{SI 4}$$

$$F_y = -i\sqrt{\frac{\beta_{sp}\gamma_N}{2}}\left(\sqrt{N+n}\,\xi_1 + \sqrt{N-n}\,\xi_2\right) \tag{SI 5}$$

In equations S1 1-5, subscripts $x$ and $y$ represent the subsidiary (orthogonally-polarised) and solitary (parallel-polarised) lasing modes of the VCSEL respectively. The field amplitudes of the subsidiary and solitary modes are represented by $E_x$ and $E_y$. The total carrier inversion between conduction and valence bands is represented by $N$, with $n$ representing the carrier inversion difference between spins of opposite polarity. $\gamma_a$ is the gain anisotropy (dichroism) rate, $\gamma_p$ is the linear birefringence rate, $\gamma_N$ is the decay rate of the carrier inversion and $\gamma_s$ is the spin-flip rate. $k$ is the field decay rate, $\alpha$ is the linewidth enhancement rate and $\mu$ is the normalized pump current (where $\mu=1$ represents the VCSEL's lasing threshold value). The injected image input, using here larger 3x3 kernel operators, is represented by $E_{inj}$ and the injection strength is controlled by $k_{inj}$. The spontaneous emission noise $F_x$ and $F_y$ are calculated using the spontaneous emission strength $\beta_{sp}$ and two independent Gaussian white noise terms, $\xi_{1,2}$, of zero mean and a unit variance. The angular frequency detuning is defined as $\Delta\omega_x = \omega_{inj}-\omega_0$, where the central frequency $\omega_0 = (\omega_x + \omega_y)/2$ lies between the frequencies of the subsidiary $\omega_x = \omega_0 + \alpha\gamma_a - \gamma_p$ and the solitary mode $\omega_y = \omega_0 + \gamma_p - \alpha\gamma_a$. $\Delta f = f_{inj}-f_x$ is the frequency detuning between the injected field and the subsidiary mode, hence $\Delta\omega_x = 2\pi\Delta f + \alpha\gamma_a - \gamma_p$. The following parameters were used to simulate the response of a VCSEL neuron: $\gamma_p$ =128 ns⁻¹, $\gamma_a$ =2 ns⁻¹, $\gamma_N$ =0.5 ns⁻¹, $\gamma_s$ =110 ns⁻¹, $\alpha$ =2, k =185 ns⁻¹, $k_{inj}$ =15 ns⁻¹ and $\beta_{sp}$ =10⁻⁵. $E_{inj}$ was introduced into the $x$ polarisation mode with a frequency detuning of $\Delta f$ = -4 GHz.

Here, image inputs were generated using the same technique as that shown in Fig.1 of the main article. Each image input implements bursts of 9 input pulses (using a 3x3 kernel), one for each Hadamard product value. The input pulses were configured with pulse widths of 100 ps and with short pulse separations of 10 ps (to fit within the temporal integration window of the VCSEL neuron, equal to ~1 ns), within a configurable 3 ns pixel duration. As before, the simulated VCSEL is responsible for the integration of inputs and the activation of spiking responses for target features. The successful integration of a 9-burst input, and the activation of a fast spiking response, is shown in the numerically calculated result of Fig. S5. Here a burst of 9 inputs, is injected into the device producing the desired spiking response from the VCSEL, hence the integration of large input bursts can be theoretically performed.

Using an image from the MNIST handwritten digit database (Fig. S6 (a)), convolution with 8 3x3 kernel operators was numerically simulated (Fig. S6 (c)-(j)). The 8 3x3 kernel operators implemented integer weights, 1 for black and -1 for white, as shown at the bottom of Fig. S6. The results showed that each kernel operator successfully detected the target features within the MNIST image. The system performed the edge detection without the activation of false positives, indicating the effective integration of the larger bursts of 9 pulses (required for operation with 3x3 kernels) by the VCSEL neuron. This result reveals that the integrate-and-fire functionality of the VCSEL neuron can be utilized for larger bursts of input pulses (within the integration time window of the device) to detect specific features in complex source images. Overall, the 8 kernel reconstruction, shown in Fig. S6(b), reveals that complete edge detection can theoretically be performed with an integrate-and-fire VCSEL neuron.

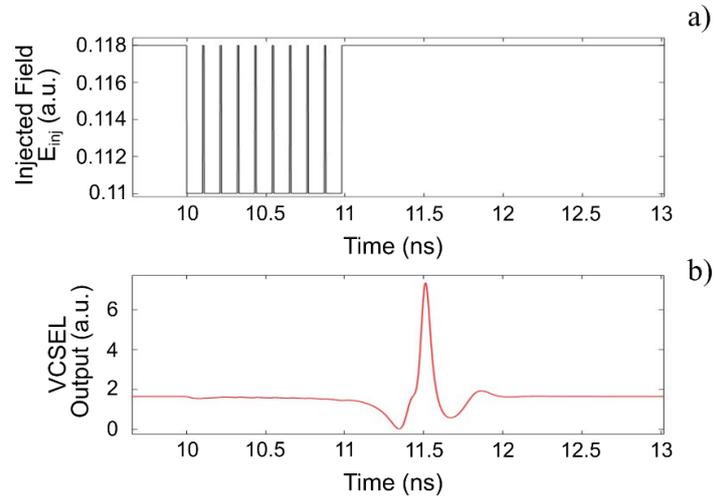

Fig. S5. Theoretical analysis of the operation of the VCSEL neuron under the injection of a large burst of 9-pulses (100 ps long and with 10 ps pulse separation). The injected field contains a burst of 9 input pulses corresponding to the detection of a target feature (a). Upon receiving this burst input, the VCSEL neuron fires a spike event (b), indicating the successful integration of the large burst and the detection of a target feature.

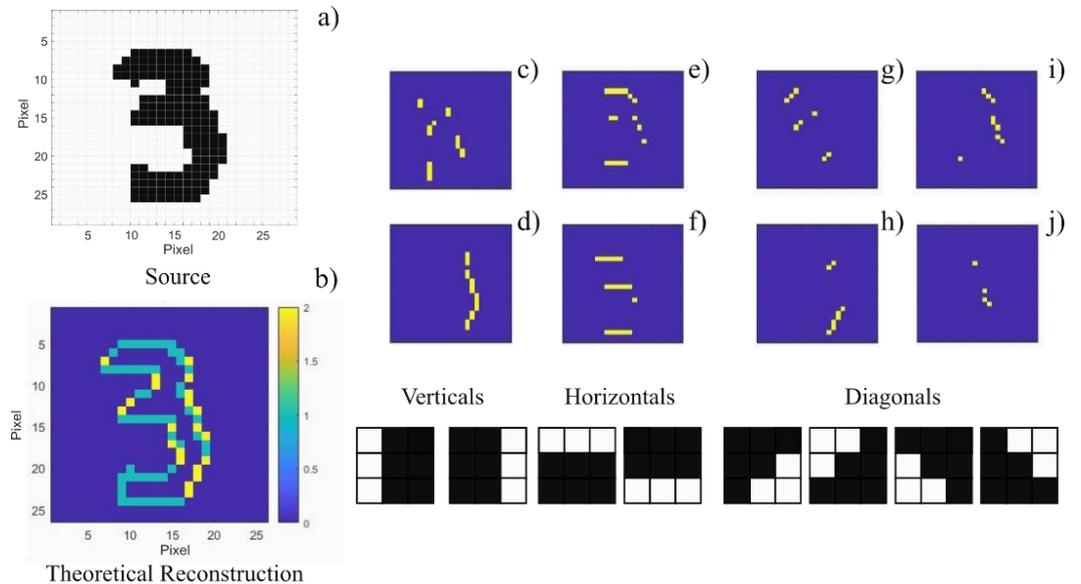

Fig. S6. Theoretical MNIST handwritten digit image edge-feature detection with a spiking VCSEL neuron using 3x3 kernel operators. The MNIST source image of a handwritten digit 3 (a) is operated on sequentially by 8 different 3x3 kernels (c)-(j). The two vertical (c)-(d), two horizontal (e)-(f), and four diagonal (g)-(j) kernel operators are shown in the insets at the bottom right side of the figure. (b) Reconstructed image combining the results obtained with the 8 kernel operators. This reveals all edges in the source image. Each plot in (b-j) is created by de-multiplexing the timeseries at the VCSEL neuron's output, plotting pixels with a positive spike activation (recognition) in yellow/grey.